%% file: ExcitedMuon.tex
\newcommand*{\ATLASLATEXPATH}{latex/}
\documentclass[cernpreprint,UKenglish,texlive=2013,txfonts]{\ATLASLATEXPATH atlasdoc}
 \pdfoutput=1

\usepackage[biblatex=true,subfigure=true,subcaption=true]{\ATLASLATEXPATH atlaspackage}
\usepackage[articletitle=false]{\ATLASLATEXPATH atlasbiblatex}

\usepackage{\ATLASLATEXPATH atlascontribute}

\RequirePackage{caption}

\usepackage{\ATLASLATEXPATH atlasphysics}
\usepackage{epstopdf}

\graphicspath{{logos/}{figures/}}

\usepackage{ExcitedMuon-defs}

\input{ExcitedMuon-metadata}

\hypersetup{pdftitle={ATLAS draft},pdfauthor={The ATLAS Collaboration}}

\begin{document}

\maketitle

\tableofcontents


\section{Introduction}
\label{sec:intro}

The Standard Model (SM) of particle physics successfully describes a wide range of
phenomena but does not explain the generational structure and mass hierarchy of quarks
and leptons.
Composite models of
fermions~\cite{Pati-Salam-1974,Pati-Salam-1975,Terazawa-1982,Eichten-1983,
Cabibbo-1985,Hagiwara-1985,Theory-baur-1990}
aim to reduce the number of matter constituents by postulating that SM fermions are
bound states of more fundamental particles.
A direct consequence of substructure would be the existence of excited fermion states.

This paper reports on a search for an excited muon $\mu^*$ using 20.3~fb$^{-1}$ of $pp$ collision
data at a centre-of-mass energy of $\sqrt{s}$ = 8~TeV recorded in 2012 with the ATLAS detector
at the Large Hadron Collider (LHC).  The search is based on a benchmark model~\cite{Theory-baur-1990}
that describes excited-fermion interactions with an effective Lagrangian containing
four-fermion contact interactions and gauge-mediated interactions.
A contact interaction decay signature not previously employed in excited leptons searches,
$\mu^* \to \mu\,j\,j$ ($j$ represents a jet), is used.

In this paper, as in ~Ref.~\cite{Theory-baur-1990}, the model is assumed to be valid
for $\mu^*$ masses up to the compositeness scale.
The contact interaction terms are described by the Lagrangian
\begin{eqnarray*}
      \mathcal{L}_{\textrm{contact}} = \frac{g_*^2}{2\Lambda^2} j^{\mu}j_{\mu},\textrm{   with }
                    j_{\mu} = \eta \overline{f}_{\textrm{L}} \gamma_{\mu} f_{\textrm{L}} 
                   +\eta^{\prime}\overline{f}^*_{\textrm{L}} \gamma_{\mu} f^*_{\textrm{L}}
                   +\eta^{\prime\prime}\overline{f}^*_{\textrm{L}} \gamma_{\mu} f_{\textrm{L}} + \textrm{h.c.},
\end{eqnarray*}
where $\Lambda$ is the compositeness scale; $j_{\mu}$ is the fermion current for ground states ($f$) and
excited states ($f^*$); $g_*$ and the $\eta$'s are constants; ``h.c.'' stands for Hermitian conjugate;
and only left-handed fermion interactions are assumed.  As is done in Ref.~\cite{Theory-baur-1990}, $g_*^2$ is set to 4$\pi$, and
$\eta$, $\eta^{\prime}$, and $\eta^{\prime\prime}$ are taken to be one for all fermions.
To calculate branching ratios, the compositeness scale $\Lambda$ is assumed to be the same
for gauge-mediated interactions, and the parameters $f$ and $f^\prime$ in Ref.~\cite{Theory-baur-1990}
are taken to be one.

\begin{figure}
\begin{center}
  \includegraphics[width=12cm]{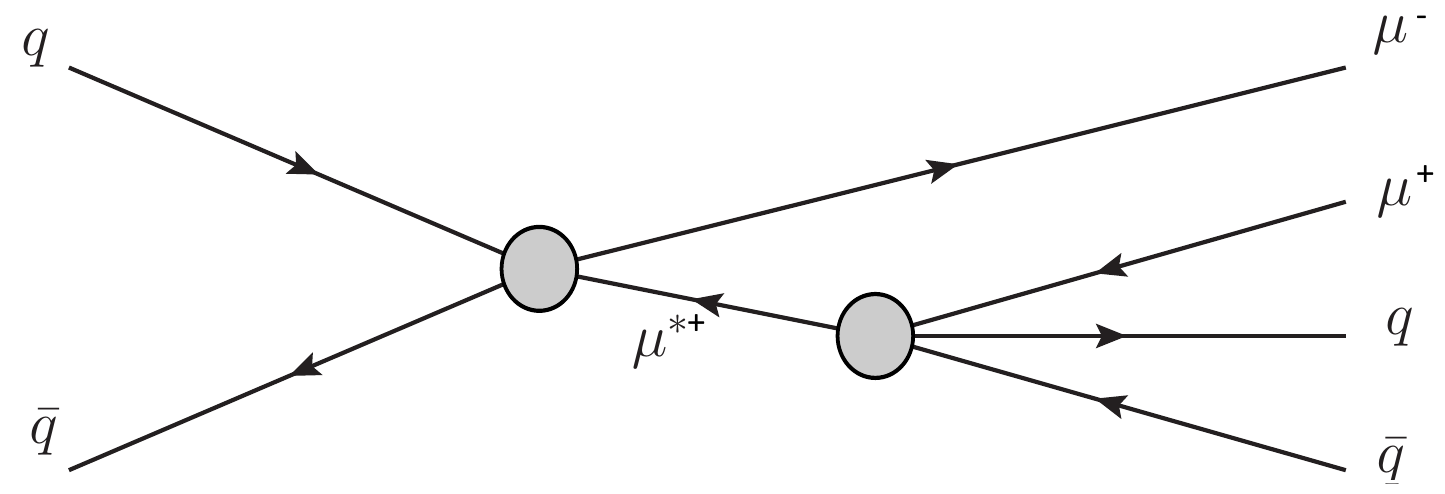}
\end{center}
\caption{Feynman diagram for the process $q\overline{q} \to \mu^* \mu \to \mu\mu q \overline{q}$, where both the production
and decay are via contact interactions. \label{fig:FeynmanDiagram}}
\end{figure}

The search described here focuses on the predominant single-$\mu^*$ production via the contact interaction
($q\overline{q} \to \mu^{*}\mu$) followed by the decay of the excited muon via the
contact interaction to $\mu q \overline{q}$,
leading to a final state with two muons and two jets
(figure~\ref{fig:FeynmanDiagram}).  
Top quarks from excited muons with masses accessible in the 8-TeV LHC data
would not have sufficient energy to form narrow jets
and are excluded from the analysis in this paper.
Previous searches at LEP~\cite{ALEPH-1996,OPAL-2000,L3-2003,DELPHI-2006},
the Tevatron~\cite{CDF-2005,CDF-2006,D0-2006,D0-2008},
and the LHC~\cite{EXOT-2011-23,EXOT-2012-21,CMS-EXO-10-016,CMS-EXO-11-034,CMS-8-TeV}
looked for the gauge-mediated decay $\mu^* \to \mu \gamma$. The analysis reported in
Ref.~\cite{CMS-8-TeV} also includes
the gauge-mediated decay $\mu* \to \mu Z$ followed by $Z \to \ell\ell$ or $q\bar{q}$.
In the model of Ref.~\cite{Theory-baur-1990}, this decay is dominant at low $\mu^*$ mass,
but for $m_{\mu^*} \gtrsim 0.25 \Lambda$,
the $\mu q \bar{q}$ decay mode is expected to have the largest branching ratio,
rising to 65\% for $m_{\mu^*} = \Lambda$.
The search reported here complements the search in the $\mu \gamma$ channel and
increases the sensitivity of the search for excited muons at higher masses.
The ATLAS Collaboration recently published~\cite{multimuon} another new search signature
for excited muons decaying via a
contact interaction to $\mu \ell \ell$, where $\ell$ is an electron or a muon.

\input{atlas-detector}

\section{Signal and background simulation}
\label{sec:simulation}

Simulation of the excited-muon signal is based on calculations from Ref.~\cite{Theory-baur-1990}.
Signal samples are generated at leading order (LO) with CompHEP 4.5.1~\cite{CompHEP}
using MSTW2008lo~\cite{MSTW} parton distribution functions (PDFs).
CompHEP is interfaced with \PYTHIA 8.170~\cite{Pythia1,Pythia2} with the AU2
parameters settings~\cite{ATL-PHYS-PUB-2012-003} for the simulation of parton showers and hadronization.
Only the production of $\mu\mu^*$ followed by the decay $\mu^* \to \mu q\bar{q}$ is simulated.
Signal samples are produced for $\Lambda$ = 5~TeV and for the $\mu^*$ masses given in table~\ref{tab:signal_region}.
The distributions of kinematic variables should be independent of $\Lambda$,
which was checked with generator-level studies.
For a compositeness scale of $\Lambda$ = 5~TeV,
cross section times branching ratios are 10.4, 2.9, and 0.21~fb for $\mu^*$ masses of 500, 1500, and
2500~GeV, respectively.
The intrinsic total width
of the $\mu^*$ is expected to be less than 8\% for $m_{\mu^*} < \Lambda$, which is smaller than the
mass resolution of about 20\% over the range of $\mu^*$ masses considered here.

The dominant background is from the process $\Zboson/\gamma^* \to \mu\mu$ produced
in association with jets ($\Zboson/\gamma^* +$ jets).  The second most important background is
\ttbar production.  Other processes, such as  diboson ($\Wboson\Wboson$, $\Wboson\Zboson$,
and $\Zboson\Zboson$), single-top, $\Wboson +$ jets, and multi-jet production, give small
contributions to the background.

The $\Zboson/\gamma^* +$ jets samples are produced by the multi-leg LO generator \SHERPA~1.4.1~\cite{Sherpa}
using CT10~\cite{CT10} PDFs.
The cross section for $\Zboson/\gamma^* \to \mu\mu$ ($m_{\mu\mu} > 70$~GeV)
plus any number of jets is 1.24~nb, calculated
at next-to-leading order (NLO), corrected by a $K$-factor~\cite{FEWZ,sigma_WZ}
to next-to-next-to-leading order (NNLO)
in QCD couplings and NLO in electroweak couplings.
The \ttbar events are generated at the parton level at NLO with \POWHEG 1.0~\cite{powheg}
and the Perugia 2011c parameter settings~\cite{Perugia2011c}, and the parton
showering is done with \PYTHIA 6.426~\cite{Pythia6}.
At least one of the $t$ or $\overline{t}$ must have a semileptonic decay ($e$, $\mu$, or $\tau$),
giving a cross section for this process of 137~pb, calculated
at NNLO + next-to-next-to-leading-log (NNLL) accuracy~\cite{sigma_ttbar}.
The diboson background samples are produced at LO by \HERWIG~6.52~\cite{Herwig}
with the AUET2 parameter settings~\cite{ATL-PHYS-PUB-2011-008} using CTEQ6L1 PDFs, and it is required
that at least one light lepton ($e$ or $\mu$) with transverse momentum (\pt) above 10~GeV be produced.
The $\Wboson +$ jets samples are produced by the multi-leg LO generator \ALPGEN~2.14~\cite{alpgen}
with \JIMMY 4.31~\cite{jimmy} and \HERWIG 6.52 using the AUET2 parameter settings with CTEQ6L1 PDFs,
and the cross section is calculated at NNLO~\cite{FEWZ,sigma_WZ}.
The multi-jet samples are generated at LO by \PYTHIA\ 8.160 using the AU2 parameter
settings with CT10 PDFs.
The single-top $t$-channel samples are generated at LO corrected to NLO + NNLL by AcerMC~3.8~\cite{AcerMC}
using the AUET2B parameters settings~\cite{ATL-PHYS-PUB-2011-014} with the CTEQ6L1 PDFs, and
the parton showering is done with \PYTHIA~6.426.
The single-top $s$- and $\Wboson t$-channel samples
are generated at NLO with MC@NLO~4.01~\cite{MCNLO1,MCNLO2,MCNLO3}
using the AUET2 parameters settings with CT10 PDFs.
The background predictions from the $\Zboson/\gamma^* +$ jets and \ttbar samples
are normalized using control regions discussed in Section~\ref{sec:bg}.
Cross sections for diboson processes are evaluated at NLO~\cite{diboson} with
an uncertainty of 5\%.
The $\Wboson +$ jets and multi-jet backgrounds are determined from the Monte Carlo (MC) samples but
are verified using data-driven methods.
A summary of the Standard Model samples used in this analysis is given in table~\ref{tab:MC}.

\begin{table}
\begin{center}
\begin{tabular} {llll} \hline \hline
    Process  & Generator & Parton showering/ & PDF  \\ 
     &  & hadronization &   \\ \hline
    \Zboson/$\gamma^*$ ($\to \mu\mu$) + jets & \SHERPA 1.4.1 & \SHERPA 1.4.1  & CT10 \\
    \ttbar ($\geq 1\ell$) & \POWHEG 1.0  & \PYTHIA 6.426 & CT10  \\
    $\Wboson\Wboson$, $\Wboson\Zboson$, $\Zboson\Zboson$ ($\geq 1\ell$)
          & \HERWIG 6.52 & \HERWIG 6.52  & CTEQ6L1  \\
    Single top, $t$-channel & AcerMC 3.8 & \PYTHIA 6.426 & CTEQ6L1 \\
    Single top, $s$-channel & MC@NLO 4.01  & \JIMMY 4.31 + \HERWIG 6.52   &  CT10  \\
    Single top, $\Wboson t$-channel & MC@NLO 4.01  & \JIMMY 4.31 + \HERWIG 6.52   &  CT10  \\
    \Wboson($\to \mu\nu$) + jets & \ALPGEN 2.14  & \JIMMY 4.31 + \HERWIG 6.52 & CTEQ6L1 \\
    Multi-jet & \PYTHIA 8.160  & \PYTHIA 8.160  & CT10  \\ \hline
    Signal ($\mu\mu^* \to \mu\mu j j$)  & CompHEP 4.5.1 & \PYTHIA 8.170
                & MSTW2008lo  \\  \hline
\end{tabular}
\end{center}
\caption{Summary of the background and signal MC sample generation used in this search.
    The columns give the process generated, the generator program,
        the parton shower program, and the PDF utilized.} \label{tab:MC}
\end{table}

The generated
samples are processed using a detailed detector simulation~\cite{SOFT-2010-01}
based on \GEANT~\hspace{-2pt}4~\cite{geant4} to propagate the particles through the detector material
and account for the detector response.
Simulated minimum-bias events are overlaid on both the signal and background samples to reproduce the effect
of additional $pp$ collisions.  The simulated events are weighted to give a distribution of the number of interactions
per bunch crossing that agrees with the data.  The simulated background and signal events are processed with the
same reconstruction programs as used for the data.

\section{Data set and event selection} \label{sec:selection}

The data were collected in 2012 during stable-beam periods of \rts = 8~TeV $pp$ collisions.  After selecting
events where the relevant parts of the detector were functioning properly, the data correspond
to an integrated luminosity of 20.3~fb$^{-1}$.  The events are required to pass at least one of two 
single-muon triggers.  The first has a nominal \pt threshold of 36~GeV, and the second has a lower
nominal threshold of 24~GeV but also has an isolation requirement that the sum of the \pt of tracks with
\pt above 1~GeV and within a distance of $\Delta R = 0.2$ of the muon, excluding the muon from the sum,
divided by the \pt of the muon is less than 0.12.

A primary vertex with at least three tracks with \pt $>$ 0.4~GeV within 200~mm
of the centre of the detector along the beam direction is required.  If there is more than
one primary vertex in an event, the one with the highest sum of $\pt^2$ is selected, where the
sum is over all tracks associated with the vertex.

Each muon candidate must be reconstructed independently in both the inner detector and the muon spectrometer.
Its momentum is determined by a combination of the two measurements using their covariance matrices.
Only muon candidates with $\pt^\mu$ above 30~GeV are considered.  Muons must have a minimum number of hits 
in the inner detector and hits in each of the inner, middle, and outer layers of the muon spectrometer.
These hit requirements, which restrict the muon acceptance to $|\eta| < 2.5$, guarantee a precise momentum
measurement.
To suppress background from cosmic rays, the muon tracks are required to have transverse and longitudinal
impact parameters $|d_0| < 0.2$~mm and $|z_0| < 1$~mm with respect to the selected primary vertex.
To reduce background from semileptonic decays of heavy-flavour hadrons, each muon is required to be isolated such that
$\sum\pt/\pt^{\mu} < 0.05$, where the sum is over inner-detector tracks with $\pt > 1$~GeV within 
a distance $\Delta R = 0.3$ of the candidate muon, excluding the muon from the sum. The muon
trigger and reconstruction efficiencies are evaluated using tag-and-probe techniques with
$\Zboson \to \mu\mu$ events~\cite{TRIG-2012-03,PERF-2014-05}, and $\pt$- and $\eta$-dependent corrections are applied
to simulated events.  Events are required to have exactly two muons of opposite charge that meet
these selection requirements.

Although electrons are not part of the signal for this search, they are used to define one of the
control regions (see Section~\ref{sec:bg}).
Each electron candidate is formed from the energy in a cluster of cells in the electromagnetic
calorimeter associated with a charged-particle track in the inner detector.  Each electron must
have \pt above 30~GeV and have $|\eta| < 2.47$ but not be in the interval $1.37 < |\eta| < 1.52$ to avoid the transition region between the barrel and endcap calorimeters.
The ATLAS tight electron identification criteria (based on the methodology described in~\cite{PERF-2013-03} and
updated for 2012 running conditions) for the transverse shower shape,
longitudinal leakage into the hadronic calorimeter, the association with an inner-detector track,
and hits in the transition radiation detector
are applied to the cluster.  An electron track is required to have transverse and longitudinal
impact parameters $|d_0| < 1$~mm and $|z_0| < 5$~mm with respect to the selected primary vertex.  Finally,
the electrons must pass the isolation requirement $\sum \ET < 0.007 \ET^e + 5$~GeV, where
the sum is of transverse energies deposited in cells within a cone of $\Delta R = 0.2$
around the electron, excluding those cells associated with the electron, and $\ET^e$ is the transverse
energy of the electron.

Jets of hadrons are reconstructed
using the anti-$k_t$ algorithm~\cite{antikt} with a radius parameter of $R = 0.4$
applied to clusters of calorimeter cells that are topologically connected.
The jets are calibrated using energy- and $\eta$-dependent correction factors derived from simulation
and with residual corrections from in-situ measurements~\cite{PERF-2011-03}.  
Jets are required to have $|\eta| < 2.8$ and $\pt > 30$~GeV. Jets that overlap ($\Delta R < 0.4$) any electron or muon
candidate satisfying the selection criteria described above are removed.  The two jets with the
highest \pt are then selected. 

The missing transverse momentum vector is the negative of the vector sum of the transverse momenta
of muons, electrons, photons~\cite{STDM-2010-08},
jets, and clusters of calibrated calorimeter cells not associated with these objects.  The missing transverse
energy is the magnitude of the missing transverse momentum vector.

\section{Background determination} \label{sec:bg}

\begin{figure}
\begin{center}
\subfigure[\Zboson/$\gamma^* +$ jets Control Region]{\includegraphics[width=8cm] {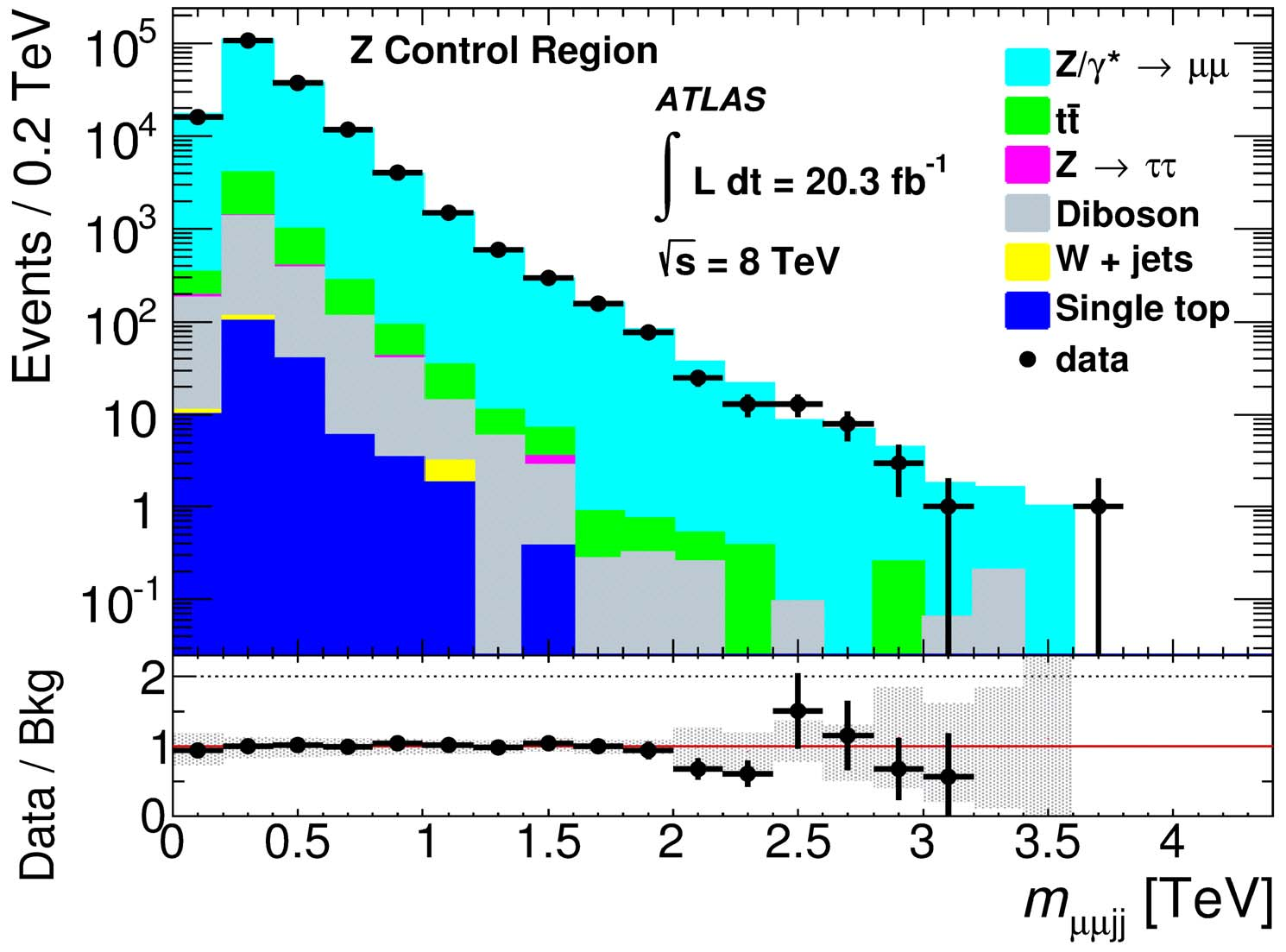}} 
\subfigure[SR 2]{\includegraphics[width=8cm]{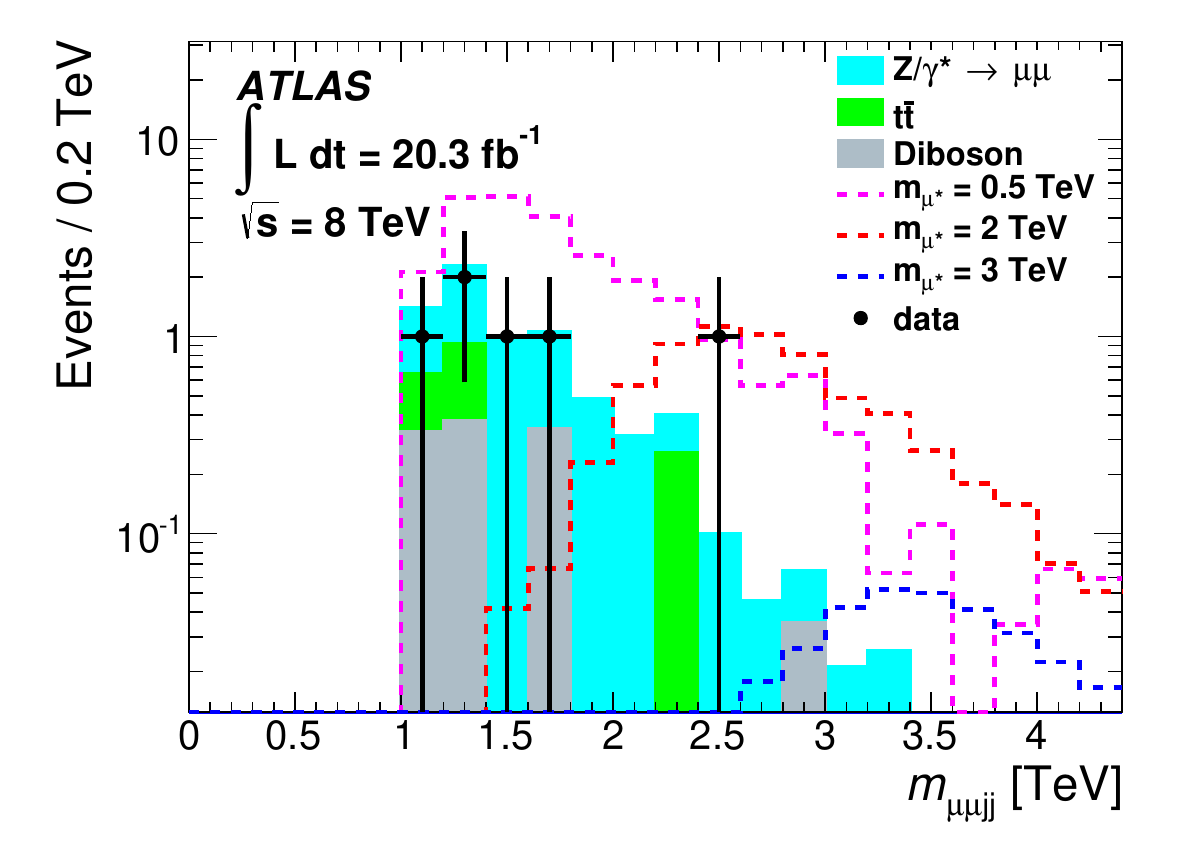}}
\end{center}
\caption{The $\mu\mu j j$ mass distribution for (a) the \Zboson/$\gamma^* +$ jets control region with the
   MC predictions and (b) for SR 2 ($m_{\mu\mu} > 550$~GeV,
   $\ST > 900$~GeV, and $m_{\mu\mu jj} > 1000$~GeV) with three representative signal distributions for
   $\mu^*$ masses of 500, 2000, and 3000~GeV and for $\Lambda$ = 5~TeV.
   The background expectations are determined after
   the fit, and the grey band on the ratio plot in (a) gives the systematic uncertainty.
   For (b) there are no expected events from the single top, $\Wboson +$ jets, and 
   $\Zboson \to \tau\tau$ processes.  SR 2 is not the
   most sensitive signal region for the latter two $m_{\mu^*}$ masses.  They are shown for comparison. \label{fig:ZCR_mmjj}}
\end{figure}

Most of the SM background contributions are estimated from the MC samples.  The expected yields
from the \Zboson/$\gamma^* +$ jets and
\ttbar production processes are normalized to the data using control regions.
The \Zboson/$\gamma^* +$ jets control region
is defined by $70 < m_{\mu\mu} < 110$~GeV in addition to the selection criteria given in Section~\ref{sec:selection}.
The \ttbar control region is defined as events that meet the selection requirements
given in Section~\ref{sec:selection},
except there is exactly one muon and one electron of opposite sign, so it should contain no signal events.
The normalization scale factors are determined from simultaneous fits to data
in the control and signal regions (see Section~\ref{sec:result}).
The scale factors are primarily determined from the control regions, giving the same values in all cases.
From the fits, the scale factor is 1.010$^{+0.087}_{-0.066}$ for the \Zboson/$\gamma^* +$ jets sample and
is 1.050~$\pm$~0.013 for the \ttbar sample.
The MC predictions agree well with the data in the control regions, as can be seen, for example, in
figure~\ref{fig:ZCR_mmjj}(a).

A jet can produce a prompt muon candidate either from the semileptonic decay of a heavy quark or from misidentification
of a charged hadron in the jet as a muon.
The expected background from jets, primarily from $\Wboson +$ jets and multi-jet processes,
is determined from MC samples, giving zero expected events.
This prediction is checked by the data-driven matrix method~\cite{TOPQ-2010-01}, which
uses isolated and non-isolated muons and their data-determined efficiencies and misidentification rates
to determine the number of prompt muons.
The matrix method predicts $-0.07 \pm 0.55$ events from these backgrounds.

\section{Signal regions}

Signal regions (SR) are defined by three kinematic variables - the dimuon invariant mass $m_{\mu\mu}$, 
the invariant mass $m_{\mu\mu j j}$ of the two muons and two jets ($j$), and \ST, the scalar sum of transverse momenta
of the four signal objects, that is,
$\ST = \pt^{\mu 1} + \pt^{\mu 2} + \pt^{j1} + \pt^{j2}$.\footnote{The $\mu j j$ invariant mass
was considered as a discriminating variable instead of one of the three selection variables.  
Several methods for selecting the correct $\mu j j$ combination
and the possibility of using both $\mu j j$
combinations were considered. No method that improved the sensitivity was found.}

For all three of these variables, the signal tends to have higher values than the backgrounds, so
all criteria are lower bounds in the selection.  The values of these bounds are chosen 
to maximize the search sensitivity for each signal mass considered
by scanning the three-dimensional parameter space for the values that minimize the expected 95\% confidence
level (CL) upper limit on the cross section times branching ratio.
The selection criteria for the signal regions
are shown in table~\ref{tab:signal_region}.  The $m_{\mu\mu j j}$ and \ST criteria
increase with increasing signal mass, but the $m_{\mu\mu}$ criterion decreases.  The latter
is because the increase in the other parameters sufficiently reduces the expected background so
that the signal efficiency may be increased by decreasing the $m_{\mu\mu}$ criterion.

The dominant background in all signal regions is from the \Zboson/$\gamma^* +$ jets process, which is
50\% of the background in SR 1, rising to 90\% or more in SR 5 through SR 10.
The \ttbar process contributes 40\% of the background in SR 1,
but this contribution falls quickly to 10\% or less in SR 3 through SR 10.
The contribution to the background from all other processes is between 10\% and 20\% in SR 1 through SR 5
and is less than 5\% for SR 6 through SR 10.

\begin{table}
\begin{center}
\begin{tabular} {rrrrrr @{.} lrrrc} \hline \hline
    \multicolumn{1}{c}{$m_{\mu^*}$} & \multicolumn{1}{c}{Signal} & \multicolumn{1}{c}{$m_{\mu\mu}$} 
      & \multicolumn{1}{c}{\ST} & $m_{\mu\mu jj}$ 
    & \multicolumn{2}{c}{Acc $\times$} & \multicolumn{1}{c}{Exp} & \multicolumn{1}{c}{Exp} & \multicolumn{1}{c}{Exp}
    & \multicolumn{1}{c}{Obs}  \\
    \hspace{1pt} [GeV]       & Region & [GeV]        &  [GeV]     
     & [GeV] & \multicolumn{2}{c}{Eff}  & \multicolumn{1}{c}{Signal}
        &  \multicolumn{1}{c}{BG} &  \multicolumn{1}{c}{BG} & \multicolumn{1}{c}{Events}  \\
      & & & & & \multicolumn{2}{c}{} & & \multicolumn{1}{c}{(prefit)} & \multicolumn{1}{c}{(postfit)} &   \\  \hline
    100 & 1\hspace{4mm} & 500 & 450 & 0 & 0&041 & 3.0$\pm$0.3 & 73$\pm$17 & 71.7$\pm$8.6 & 71 \\
    300 & 2\hspace{4mm} & 550 & 900 & 1000 & 0&088 & 12.5$\pm$0.9 & 10.1$\pm$3.5 & 7.8$\pm$2.2 &  6 \\
    500 & 2\hspace{4mm} & 550 & 900 & 1000 & 0&15 & 29.4$\pm$1.6 & 10.1$\pm$3.5 & 7.8$\pm$2.2 &  6   \\
    750 & 3\hspace{4mm} & 450 & 900 & 1300 & 0&23 & 43.7$\pm$2.2 & 6.6$\pm$2.5 & 5.8$\pm$1.9 &  5   \\
    1000 & 4\hspace{4mm} & 450 & 1050 & 1300 & 0&31 & 44.1$\pm$1.8 & 4.6$\pm$1.8 & 4.6$\pm$1.9 &  5   \\
    1250 & 5\hspace{4mm} & 450 & 1200 & 1500 & 0&38 & 29.8$\pm$1.3 & 2.1 $\pm$0.9 & 2.1$\pm$1.0 &  3   \\
    1500 & 6\hspace{4mm} & 400 & 1200 & 1700 & 0&38 & 19.8$\pm$0.8 & 1.5$\pm$0.7 & 1.3$\pm$0.6 &  1   \\
    1750 & 7\hspace{4mm} & 300 & 1350 & 1900 & 0&41 & 11.8$\pm$0.5 & 0.9$\pm$0.7  & 0.9$\pm$0.5 &  2   \\
    2000 & 8\hspace{4mm} & 300 & 1350 & 2000 & 0&40 & 6.6$\pm$0.2 & 0.7$\pm$0.4 & 0.7$\pm$0.4 &  2   \\
    2250 & 9\hspace{4mm} & 300 & 1500 & 2100 & 0&37 & 3.4$\pm$0.1 & 0.4$\pm$0.3 & 0.4$\pm$0.3 &  2   \\
    2500 & 10\hspace{4mm} & 110 & 1650 & 2300 & 0&39 & 1.65$\pm$0.07 & 0.2$^{+1.3}_{-0.2}$ & 0.2$^{+1.0}_{-0.2}$ &  2   \\
    2750 & 10\hspace{4mm} & 110 & 1650 & 2300 & 0&45 &  0.72$\pm$0.02 & 0.2$^{+1.3}_{-0.2}$  & 0.2$^{+1.0}_{-0.2}$ &  2    \\
    2900 & 10\hspace{4mm} & 110 & 1650 & 2300 & 0&45 & 0.52$\pm$0.02 & 0.2$^{+1.3}_{-0.2}$  & 0.2$^{+1.0}_{-0.2}$ & 2    \\
    3000 & 10\hspace{4mm} & 110 & 1650 & 2300 & 0&46 &  0.38$\pm$0.02 & 0.2$^{+1.3}_{-0.2}$  & 0.2$^{+1.0}_{-0.2}$ &  2    \\
    3100 & 10\hspace{4mm} & 110 & 1650 & 2300 & 0&45 & 0.30$\pm$0.02 & 0.2$^{+1.3}_{-0.2}$  & 0.2$^{+1.0}_{-0.2}$ &  2   \\  \hline
\end{tabular}
\end{center}
\caption{The signal masses considered and the corresponding signal regions are listed.
The $m_{\mu\mu}$, \ST, and $m_{\mu\mu jj}$
values giving the lower bound of each signal region are listed, along with the acceptance times efficiency,
the expected number of signal events
($\Lambda = 5$~TeV), expected number of background events before and after the fit discussed in Section~\ref{sec:result},
and the number of events observed in the data.
The uncertainties in the expected numbers of signal and background events are the systematic uncertainties.
The numbers of events observed are discussed in Section~\ref{sec:result}.} \label{tab:signal_region}
\end{table}

\section{Systematic uncertainties}

Contributions to the systematic uncertainties in the background and signal yield predictions
stem from both experimental and theoretical sources, as discussed below.

The luminosity is derived using the methodology in Ref.~\cite{DAPR-2011-01}
and has an uncertainty of 2.8\%.  The luminosity uncertainty for the backgrounds
is less than this because the largest backgrounds (\Zboson/$\gamma^* +$ jets and \ttbar) are
normalized using control regions.

Uncertainties in the MC modelling of the detector, particularly for muons and jets in this
analysis, must be taken into account and are derived from detailed studies of data.
One-standard-deviation variation of a given parameter is determined, and
then the parameter is varied up and down in the simulation by this amount to determine
the effect on the signal and background yields.

The uncertainty in the jet energy scale is the largest contribution to the systematic uncertainty
in the signal yield and a significant contribution to the uncertainty in the backgrounds.  
The uncertainty in the jet energy resolution also makes a contribution.  These uncertainties
are determined from \pt balance in $\gamma +$ jet and $\Zboson +$ jet events
and in events with high-\pt jets recoiling against multiple, low-\pt jets~\cite{PERF-2011-03,PERF-2012-01}.  
The uncertainty in contributions from additional energy deposited in the
calorimeters from other $pp$ interactions in the event is also included.  The various effects are investigated
separately and combined to give the values summarized in tables~\ref{tab:syst_sig} and~\ref{tab:syst_bg}.

Muon performance is determined in $\Zboson \to \mu \mu$ events.  The most important
parameters for this analysis are the muon efficiency and the muon spectrometer \pt resolution.
The inner-detector resolution and the muon \pt scale are found to have negligible effect.
The uncertainty in the trigger efficiency is less than 2\% for the backgrounds and less than 1\%
for the signal yield.

The uncertainties in the signal and background yield predictions due to uncertainties in PDFs
have two contributions.  The first is from one-standard-deviation variation of the parameters
of the relevant PDFs (Section~\ref{sec:simulation}).  The second is a comparison
with the alternative NNPDF2.1 PDF set~\cite{NNPDF}. 
These variations produce changes in the predicted cross section and in kinematical distributions,
which in turn affect the acceptance times efficiency.
For the background, both effects are included in the systematic uncertainty.
For the signal yield, the uncertainty in the acceptance times efficiency is included, but the uncertainty
in the cross section is considered part of the uncertainty in the theoretical prediction and is
not included in the statistical analysis.

The uncertainty in the background modelling in the signal regions is estimated by examining
how well the MC prediction agrees with the data in two validation regions selected to
be similar in kinematics to the signal regions but containing no signal.  Both validation
regions require the same selection as the signal regions except that $m_{\mu\mu jj} < 500$~GeV
and $m_{\mu\mu} > 200$~GeV with no selection on $S_\textrm{T}$.  Requiring the missing
transverse energy be greater (less) than 50~GeV (40~GeV) selects a validation
region dominated by \ttbar ($Z/\gamma^*$) events.    
For some of the kinematic variables, an extrapolation of the predicted yield
from the validation regions to the signal regions is necessary to evaluate
possible mismodelling effects.
Of the several kinematic variables studied, only the modelling of the \ST variable
is found to have a significant effect.
A linear fit to the ratio of the number of data events to the MC expectation
is extrapolated to higher values of \ST, and
the deviation from unity symmetrized about zero
gives the uncertainty, referred to as ``\Zboson/$\gamma^* +$ jets modelling'' and ``\ttbar
modelling'' in table~\ref{tab:syst_bg}.  For both validation regions,
the linear fit is consistent within the statistical uncertainties
with a flat line at a ratio of one.

To produce sufficient numbers of events for high dimuon masses, the $\Zboson/\gamma^*$ MC samples
were produced in bins of dimuon mass above the \Zboson mass.  For the
\ST and $m_{\mu\mu j j}$ criteria in this analysis,
this yields zero events in SR 7 through SR 10 for some ranges
of the $m_{\mu\mu}$ distribution (for example, 110 to 400 GeV for SR 10).
For these signal regions, an additional systematic uncertainty (referred to as
``\Zboson/$\gamma^* +$ jets extrapolation'' in table~\ref{tab:syst_bg})
is estimated by loosening the $S_\textrm{T}$ criteria and extrapolating into the signal region.
The uncertainty introduced by this procedure is small except in SR 10, where the
effect on the statistical analysis is still small because the predicted number of background events is only 0.2.  

Additional sources of uncertainty in the acceptance times efficiency are initial-state radiation,
final-state radiation, renormalization and factorization scales, and the beam energy.  The effects of initial- and
final-state radiation are determined in generator-level studies by varying the relevant \PYTHIA parameters
and are less than 1\%.
The effect of the beam energy uncertainty (0.65\%)~\cite{CERN-ATS-2013-040} is determined by varying the momentum
fraction of the initial partons in the PDFs by this amount, giving a change of less than 1\%.
The renormalization and factorization scales are independently varied in the simulation by factors of 2 and 1/2,
changing the expected signal acceptance times efficiency by about 2\% at low mass and by less than 1\% for masses
above 750~GeV.

The uncertainties in the signal yield depend on the $\mu^*$ mass, and the largest contributions are
summarized in table~\ref{tab:syst_sig} for three representative masses.
For the signal yield,
uncertainties in jet energy scale, PDFs, and luminosity are the dominant sources.
The uncertainties in the background depend on the signal region,
and the largest contributions are shown in table~\ref{tab:syst_bg} for
three representative regions.
The most significant contributions to the
background uncertainty are from the modelling of the \Zboson/$\gamma^* +$ jets
and \ttbar processes.  The jet energy scale and the parton
distribution functions also make significant contributions.
Any source of systematic uncertainty contributing less than 2\% to the background
for all signal regions and less than 1\% to the signal yield for all $\mu^*$ masses
would have negligible effect in the statistical analysis in Section~\ref{sec:result} and is not included.

\begin{table}
\begin{center}
\begin{tabular} {lr @{.} l r @{.} l r @{.} l} \hline
     $m_{\mu^*}$ [GeV]  & \multicolumn{2}{c}{500} & \multicolumn{2}{c}{1500} & \multicolumn{2}{c}{2500}  \\ \hline
     Luminosity & 2&8 & 2&8 & 2&8 \\
     Jet energy scale & 2&7 & 1&5 & 1&1  \\
     Hadronization and factorization scales & 2&0 & 0&5 & 0&1 \\
     PDFs & 3&0 & 2&5 & 2&7 \\
     Muon efficiency & 0&7 & 0&8 & 0&9  \\
     Jet energy resolution & 0&9 & 0&2 & 0&5  \\
     Muon spectrometer resolution & 0&3 & $<$0&1 & 0&2  \\ \hline
     Total & 5&1 & 4&1 & 4&2 \\ \hline
\end{tabular}
\end{center}
\caption{Largest contributions to the relative systematic uncertainty in the signal yield.
     The uncertainties for the hadronization and factorization scales and for the PDFs are only
     for the signal acceptance times efficiency.
     All uncertainties are given in percent and are determined after the fit discussed in Section~\ref{sec:result}.
     } \label{tab:syst_sig}
\end{table}

\begin{table}
\begin{center}
\begin{tabular} {lr @{.} l r @{.} l r @{.} l} \hline
     Signal region & \multicolumn{2}{c}{2} & \multicolumn{2}{c}{6} & \multicolumn{2}{c}{10} \\ \hline
    \Zboson/$\gamma^* +$ jets modelling & \multicolumn{2}{l}{25} & \multicolumn{2}{l}{$\,\,$47} & \multicolumn{2}{l}{$\,\,$65}  \\
    Jet energy scale & \multicolumn{2}{l}{19} & 9&0 & 6&2 \\
    \ttbar modelling & \multicolumn{2}{l}{12} & $<$0&1 & $\,<$0&1 \\
    Muon spectrometer resolution & $\,\,\,$6&2 & 0&6 & \multicolumn{2}{l}{$\,\,$63} \\
    PDFs & 4&2 & 8&8 & \multicolumn{2}{l}{$\,\,$17} \\
    Jet energy resolution & 3&2 & 1&7 & 0&6  \\
    Muon efficiency & 0&7 & 0&8 & 0&9  \\
    Luminosity & 0&4 & 0&1 & $<$0&1 \\
    \Zboson/$\gamma^* +$ jets extrapolation & \multicolumn{2}{c}{--}  & \multicolumn{2}{c}{--}  & \multicolumn{2}{l}{500} \\  \hline
    Total & \multicolumn{2}{l}{35} & \multicolumn{2}{l}{$\,\,$49} & \multicolumn{2}{l}{500}  \\ \hline 
\end{tabular}
\end{center}
\caption{Largest contributions to the relative systematic uncertainty in the expected background
     for three representative signal regions.
     All uncertainties are given in percent and are determined after the fit discussed in Section~\ref{sec:result}.
     } \label{tab:syst_bg}
\end{table}

\section{Results}
\label{sec:result}

For each $\mu^*$ mass considered, the numbers of events in the corresponding signal region
and in the two control regions
are simultaneously fit~\cite{histfit} using a profile likelihood method~\cite{Profile,ProfileErratum}.
The likelihood function models the number of events as a Poisson distribution
and the systematic effects are modelled using nuisance parameters with lognormal constraints.
The parameters of interest in the fit are the signal yield in the corresponding signal region and
the normalizations of the $Z/\gamma^*$ and \ttbar backgrounds, with the
latter two being primarily determined in the fit by the events in the control regions.
The possible contribution of signal to the control regions is included in the fit and found
to be negligible.
Correlations of the systematic uncertainties are taken into account.

As an example of the result of the fit,
the $m_{\mu\mu jj}$ distribution for signal region 2 is shown in figure~\ref{fig:ZCR_mmjj}(b) for the data, expected backgrounds,
and three signal predictions for $\Lambda$ = 5~TeV (the signal regions for the higher masses
have fewer background events).  The expected and observed numbers of events for each signal mass
considered are shown in table~\ref{tab:signal_region} for $\Lambda$ = 5~TeV.
Due to correlations among the nuisance parameters, the uncertainties
on the expected backgrounds are reduced after the fits.
The data are consistent with the
Standard Model expectations, and no significant excess is observed.  Thus, limits 
on the cross section times branching ratio as a function of the $\mu^*$ mass are calculated.

A modified frequentist $CL_s$ method~\cite{CLS1,CLS2}
is used to derive the 95\% CL upper limits on the signal yield.
The expected limit is the median limit for a large number of background-only pseudo-experiments.
The one- and two-standard-deviations bands cover 68\% and 95\%, respectively, of the pseudo-experiment limits.
The observed limit is the 95\% CL limit
for the observed number of events.
The $p$-value is a measure of how well the background-only hypothesis models the data.
For a signal region, it is the fraction of background-only pseudo-experiments where the fitted signal value
is greater than that for the observed data.

The smallest $p$-values
are for SR 9 and 10 with values of 0.034 and 0.099, respectively, corresponding to 1.8 and 1.3 standard deviations
on one side of a Gaussian distribution.
Some kinematic properties
of the events in these signal regions are given in table~\ref{tab:events}.
There is one event (event A) that is in all signal regions.

\begin{table}
\begin{center}
\begin{tabular} {rrrrrrrrrrr} \hline
      &    & \multicolumn{1}{c}{$m_{\mu\mu}$} & $m_{\mu\mu jj}$ & \multicolumn{1}{c}{\ST} &
     \multicolumn{1}{c}{$m_{\mu_1 jj}$} & \multicolumn{1}{c}{$m_{\mu_2 jj}$} & $\pt^{\mu_1}$ & $\pt^{\mu_2}$ 
      & $\pt^{j_1}$ & $\pt^{j_2}$ \\
     \multicolumn{1}{c}{Event} & \multicolumn{1}{c}{SR} &   [GeV]      &
     [GeV]       &  [GeV]&    [GeV]      &      [GeV]         &   [GeV]  &   [GeV]  &   [GeV] &   [GeV]  \\ \hline
    A & all & 1800 & 2410 & 1820 & 1200 & 1090 & 650 & 630 & 350 & 190 \\ 
    B & 7--9 & 310 & 2250 & 2010 & 2200 & 630 & 840 & 46 & 990 & 130  \\
    C & 10 & 113 & 2440 & 1760 & 2230 & 1850 & 150 & 35 & 890 & 690  \\ \hline
\end{tabular}
\end{center}
\caption{Values of $\mu\mu$ mass, $\mu\mu jj$ mass, \ST, $\mu jj$ mass for each
$\mu jj$ combination, and \pt of each muon and jet
for the three events in SR 9 or 10.
} \label{tab:events}
\end{table}

\begin{figure}
\begin{center}
  \includegraphics[width=12cm]{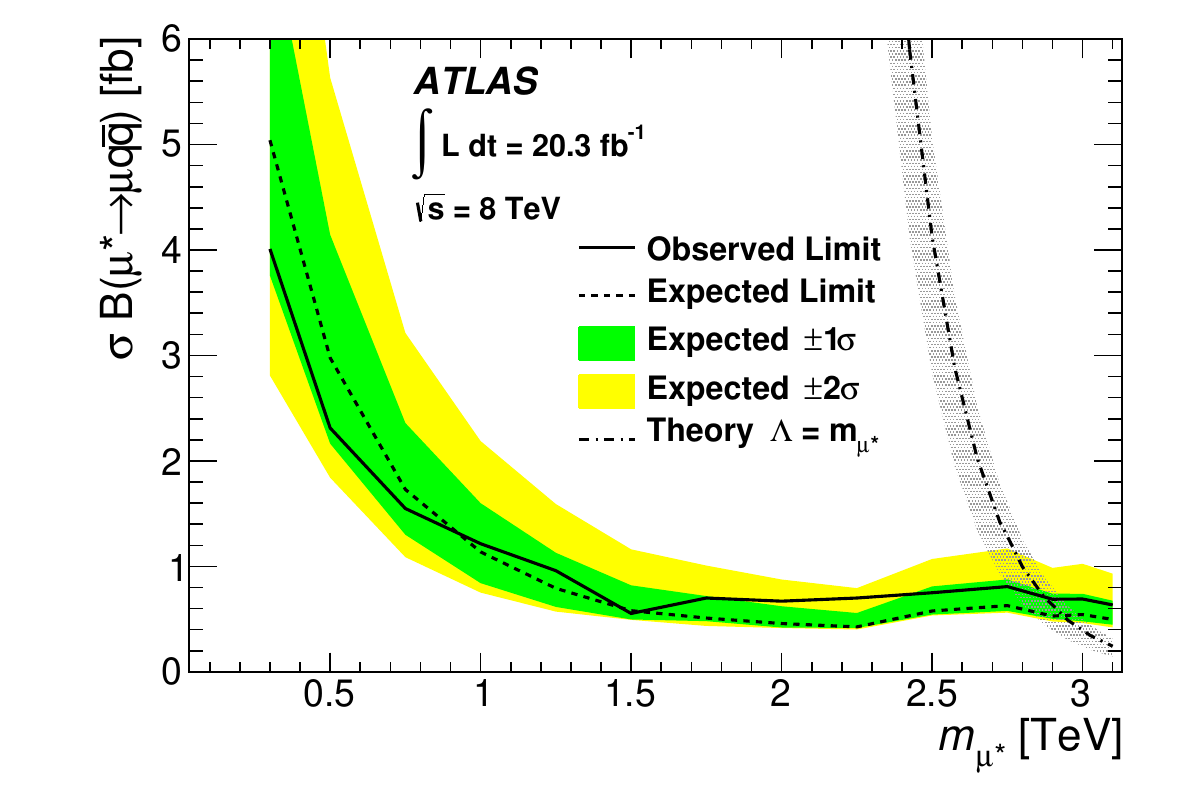}
\end{center}
\caption{Limit at 95\% CL on cross section times branching ratio
   $\sigma (pp \to \mu \mu^*) B(\mu^* \to \mu q \bar{q})$ as a function of the $\mu^*$ mass,
   where $q$ is any quark except the top quark.
   The solid line is the limit and the dotted line is the expected limit.
   The theoretical $\sigma B$ for the limiting case $\Lambda = m_{\mu^*}$ along with
   its uncertainties is also shown.} \label{fig:sigmaB}
\end{figure}

An upper limit on the cross section times branching ratio
$\sigma (pp \to \mu \mu^*) \times B(\mu^* \to \mu q\bar{q})$ (figure~\ref{fig:sigmaB}) is determined
for each signal mass from the limit on the signal yield at the 95\% CL.
The theoretical uncertainties are not included in either the $\sigma B$ or $\Lambda$ limit
determinations.
For $m_{\mu^*}$ above 1.3~TeV, the limit is between 0.6 and 1~fb.
The theoretical expectation for $\Lambda = m_{\mu^*}$ is also shown.
The theoretical band represents uncertainties from PDFs and from renormalization and factorization scales.

\begin{figure}
\begin{center}
  \includegraphics[width=12cm]{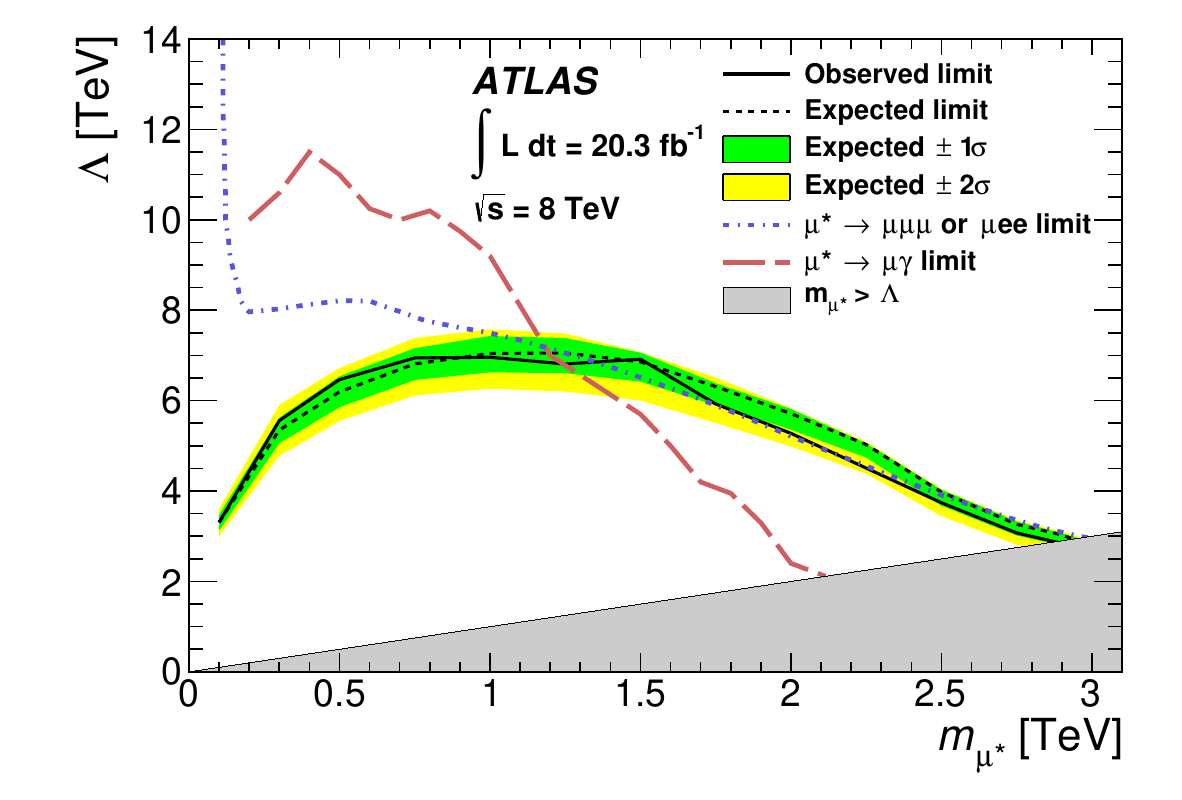}
\end{center}
\caption{Limit at 95\% CL on the compositeness scale $\Lambda$ as a function
   of the $\mu^*$ mass.  The solid line is the observed limit and the short dashed
   line is the expected limit.  Also indicated are previous results from ATLAS
   based on $\mu^* \to \mu \gamma$ (long dashed line) and $\mu^* \to \mu \ell \ell$
   (dot-dashed line), where $\ell$ is an electron or muon.}
   \label{fig:Lambda_bound}
\end{figure}

The expected cross section and branching ratio depend on the $\mu^*$ mass and on $\Lambda$~\cite{Theory-baur-1990}.
For each signal mass, the limit on $\sigma B$ is translated into a lower bound on
the compositeness scale (figure~\ref{fig:Lambda_bound}).
The bound is the value of $\Lambda$ for which the theoretical
prediction of $\sigma B(m_{\mu^*}, \Lambda)$ is equal to the upper limit on $\sigma B$.
The region with $m_{\mu^*} > \Lambda$ is unphysical.
For the limiting case where
$\Lambda = m_{\mu^*}$, excited-muon masses below 2.9~TeV are excluded.
Previous limits set by ATLAS~\cite{EXOT-2012-21,multimuon} 
are also shown.  for masses above 1200~GeV,
the analysis presented here improves upon the limits from $\mu^* \to \mu \gamma$
and is comparable to those from $\mu^* \to \mu\ell\ell$.

\FloatBarrier

\section{Conclusion}
\label{sec:conclusion}

The results of a search for excited muons decaying to $\mu j j$ via a contact
interaction are reported based
on data from \rts = 8~TeV $pp$ collisions collected with the ATLAS detector at the LHC
corresponding to an integrated luminosity of 20.3~fb$^{-1}$.  The observed data are
consistent with SM expectations.  An upper limit is set at 95\% CL on the cross section
times branching ratio $\sigma B(\mu^* \to \mu q\bar{q})$ as a function of the
excited-muon mass.  For $m_{\mu^*}$ between 1.3 and 3.0~TeV, the limit on $\sigma B$ is between 0.6 and 1~fb.


The $\sigma B$
upper limits are converted to lower bounds on the compositeness scale $\Lambda$.
In the limiting case where $\Lambda = m_{\mu^*}$, excited-muon masses below
2.9~TeV are excluded.
At higher $\mu^*$ masses, the signature explored in this paper, $\mu* \to \mu\, j\, j$, has better
sensitivity than the traditional signature $\mu^* \to \mu\,\, \gamma$.  For $\mu^*$ masses
above 1.2~TeV, the sensitivity is comparable to 
a previous search using the signature $\mu^* \to \mu \ell\ell$.
In models other than the benchmark model used here, the branching ratios
to these modes could be different, affecting their relative importance for limits
on the compositeness scale.

\section*{Acknowledgements}

\input{acknowledgements/Acknowledgements}

\clearpage


\printbibliography

\newpage \input{atlas_authlist}



\end{document}

%% file: ExcitedMuon-metadata.tex
\AtlasTitle{A search for an excited muon decaying to a muon and two jets in $pp$
collisions at $\sqrt{s}$ = 8~TeV with the ATLAS detector}

\author{The ATLAS Collaboration}

\AtlasRefCode{EXOT-2015-01}

 \AtlasNote{ATL-COM-PHYS-2015-414}

 \PreprintIdNumber{CERN-PH-EP-2015-298}

\AtlasJournal{New Journal of Physics}
\AtlasJournalRef{New J. Phys. 18 (2016) no. 7, 073021}
\AtlasDOI{10.1088/1367-2630/18/7/073021}

\AtlasAbstract{A new search signature for excited leptons is explored.
Excited muons are sought in the channel $pp \to \mu\mu^* \to \mu \mu\textrm{ jet jet}$,
assuming both the production and decay occur via a contact interaction.  The analysis
is based on 20.3~fb$^{-1}$ of $pp$ collision data at a centre-of-mass energy of \rts = 8~TeV
taken with the ATLAS detector at the Large Hadron Collider.  No evidence of excited muons is found, and
limits are set at the 95\% confidence level on the cross section times branching ratio as a
function of the excited-muon mass $m_{\mu^*}$.
For $m_{\mu^*}$ between 1.3~TeV and 3.0~TeV,
the upper limit on $\sigma B(\mu^* \to \mu q \bar{q}$) is between 0.6 and 1~fb.
Limits on $\sigma B$ are converted to lower bounds on the compositeness scale $\Lambda$.
In the limiting case $\Lambda = m_{\mu^*}$, excited muons with a mass below 2.9~TeV are excluded.
With the same model assumptions,
these limits at larger $\mu^*$ masses improve upon previous limits from traditional searches based on the
gauge-mediated decay $\mu^* \to \mu \gamma$.
}

\AtlasCoverSupportingNote{Search for excited muons decaying via a contact interaction}{https://cds.cern.ch/record/1971084}
\AtlasCoverCommentsDeadline{Wednesday, April 6, 2016}

\AtlasCoverAnalysisTeam{Glenn Amundsen, Craig Blocker}

\AtlasCoverEdBoardMember{Stephane~Willocq~(chair), Masahiro~Kuze, Iacopo~Vivarelli, Haichen~Wang}

\AtlasCoverEgroupEditors{atlas-exot-2015-01-editors@cern.ch}

\AtlasCoverEgroupEdBoard{atlas-exot-2015-01-editorial-board@cern.ch}

%% file: atlas-detector.tex
\newcommand{\AtlasCoordFootnote}{%
ATLAS uses a right-handed coordinate system with its origin at the nominal interaction point (IP)
in the centre of the detector and the $z$-axis along the beam pipe.
The $x$-axis points from the IP to the centre of the LHC ring,
and the $y$-axis points upwards.
Cylindrical coordinates $(r,\phi)$ are used in the transverse plane, 
$\phi$ being the azimuthal angle around the $z$-axis.
The pseudorapidity is defined in terms of the polar angle $\theta$ as $\eta = -\ln \tan(\theta/2)$.
Angular distance is measured in terms of $\Delta R \equiv \sqrt{(\Delta\eta)^{2} + (\Delta\phi)^{2}}$.}

\section{ATLAS detector}
\label{sec:atlas1}

The ATLAS experiment~\cite{PERF-2007-01} uses a multi-purpose particle detector
with a forward-backward symmetric cylindrical geometry and a near $4\pi$ coverage in 
solid angle.\footnote{\AtlasCoordFootnote}
It consists of an inner tracking detector surrounded by a thin superconducting solenoid
providing a \SI{2}{\tesla} axial magnetic field, electromagnetic and hadron calorimeters, and a muon spectrometer.
The inner tracking detector covers the pseudorapidity range $|\eta| < 2.5$.
It consists of silicon pixel, silicon microstrip, and transition radiation tracking detectors.
Lead/liquid-argon (LAr) sampling calorimeters provide electromagnetic (EM) energy measurements
with high granularity.
A hadronic steel/scintillator-tile calorimeter covers the central pseudorapidity range ($|\eta| < 1.7$).
The endcap and forward regions are instrumented with LAr calorimeters
for EM and hadronic energy measurements up to $|\eta| = 4.9$.
The muon spectrometer surrounding the calorimeters covers the
pseudorapidty range $|\eta| < 2.7$ and is based on
three large air-core toroid superconducting magnets with eight coils each.
Their bending power is in the range from \num{2.0} to \SI{7.5}{\tesla\metre}.
The muon spectrometer consists of three stations of precision tracking chambers
and fast detectors for triggering.
The majority of the precision tracking chambers are composed of drift tubes, while cathode-strip
chambers provide coverage in the inner stations of the forward region for 2.0 $< |\eta| <$ 2.7.
A three-level trigger system is used to select events.
The first-level trigger is implemented in hardware and uses a subset of the detector information
to reduce the accepted rate to at most \SI{75}{\kilo\hertz}.
This is followed by two software-based trigger levels that
together reduce the accepted event rate to \SI{400}{\hertz} on average,
depending on the data-taking conditions during 2012.

%% file: acknowledgements/Acknowledgements.tex

We thank CERN for the very successful operation of the LHC, as well as the
support staff from our institutions without whom ATLAS could not be
operated efficiently.

We acknowledge the support of ANPCyT, Argentina; YerPhI, Armenia; ARC, Australia; BMWFW and FWF, Austria; ANAS, Azerbaijan; SSTC, Belarus; CNPq and FAPESP, Brazil; NSERC, NRC and CFI, Canada; CERN; CONICYT, Chile; CAS, MOST and NSFC, China; COLCIENCIAS, Colombia; MSMT CR, MPO CR and VSC CR, Czech Republic; DNRF and DNSRC, Denmark; IN2P3-CNRS, CEA-DSM/IRFU, France; GNSF, Georgia; BMBF, HGF, and MPG, Germany; GSRT, Greece; RGC, Hong Kong SAR, China; ISF, I-CORE and Benoziyo Center, Israel; INFN, Italy; MEXT and JSPS, Japan; CNRST, Morocco; FOM and NWO, Netherlands; RCN, Norway; MNiSW and NCN, Poland; FCT, Portugal; MNE/IFA, Romania; MES of Russia and NRC KI, Russian Federation; JINR; MESTD, Serbia; MSSR, Slovakia; ARRS and MIZ\v{S}, Slovenia; DST/NRF, South Africa; MINECO, Spain; SRC and Wallenberg Foundation, Sweden; SERI, SNSF and Cantons of Bern and Geneva, Switzerland; MOST, Taiwan; TAEK, Turkey; STFC, United Kingdom; DOE and NSF, United States of America. In addition, individual groups and members have received support from BCKDF, the Canada Council, CANARIE, CRC, Compute Canada, FQRNT, and the Ontario Innovation Trust, Canada; EPLANET, ERC, FP7, Horizon 2020 and Marie Sk{\l}odowska-Curie Actions, European Union; Investissements d'Avenir Labex and Idex, ANR, R{\'e}gion Auvergne and Fondation Partager le Savoir, France; DFG and AvH Foundation, Germany; Herakleitos, Thales and Aristeia programmes co-financed by EU-ESF and the Greek NSRF; BSF, GIF and Minerva, Israel; BRF, Norway; Generalitat de Catalunya, Generalitat Valenciana, Spain; the Royal Society and Leverhulme Trust, United Kingdom.

The crucial computing support from all WLCG partners is acknowledged
gratefully, in particular from CERN and the ATLAS Tier-1 facilities at
TRIUMF (Canada), NDGF (Denmark, Norway, Sweden), CC-IN2P3 (France),
KIT/GridKA (Germany), INFN-CNAF (Italy), NL-T1 (Netherlands), PIC (Spain),
ASGC (Taiwan), RAL (UK) and BNL (USA) and in the Tier-2 facilities
worldwide.

%% file: atlas_authlist.tex
 
\begin{flushleft}
{\Large The ATLAS Collaboration}

\bigskip

G.~Aad$^\textrm{\scriptsize 100}$,    
B.~Abbott$^\textrm{\scriptsize 127}$,    
J.~Abdallah$^\textrm{\scriptsize 76}$,    
O.~Abdinov$^\textrm{\scriptsize 13,*}$,    
B.~Abeloos$^\textrm{\scriptsize 131}$,    
R.~Aben$^\textrm{\scriptsize 118}$,    
M.~Abolins$^\textrm{\scriptsize 105}$,    
O.S.~AbouZeid$^\textrm{\scriptsize 165}$,    
H.~Abramowicz$^\textrm{\scriptsize 160}$,    
H.~Abreu$^\textrm{\scriptsize 159}$
% ...
\bigskip
\\

% ...
$^{13}$Institute of Physics, Azerbaijan Academy of Sciences, Baku; Azerbaijan.\\
% ...
$^{76}$University of Iowa, Iowa City IA; United States of America.\\
% ...
$^{100}$CPPM, Aix-Marseille Universit\'e, CNRS/IN2P3, Marseille; France.\\
% ...
$^{105}$Department of Physics and Astronomy, Michigan State University, East Lansing MI; United States of America.\\
% ...
$^{118}$Nikhef National Institute for Subatomic Physics and University of Amsterdam, Amsterdam; Netherlands.\\
% ...
$^{127}$Homer L. Dodge Department of Physics and Astronomy, University of Oklahoma, Norman OK; United States of America.\\
% ...
$^{131}$LAL, Universit\'e Paris-Sud, CNRS/IN2P3, Universit\'e Paris-Saclay, Orsay; France.\\
% ...
$^{159}$Department of Physics, Technion, Israel Institute of Technology, Haifa; Israel.\\
$^{160}$Raymond and Beverly Sackler School of Physics and Astronomy, Tel Aviv University, Tel Aviv; Israel.\\
% ...
$^{165}$Department of Physics, University of Toronto, Toronto ON; Canada.\\
% ...
$^{*}$ Deceased

\end{flushleft}
